# Phonon-polaritonic skyrmions: Transition from bubble- to Néel-type

Florian Mangold[†,1], Enrico Baù[†,2], Lin Nan[2], Julian Schwab[1], Thorsten Gölz[2], Andrea Mancini[2], Bettina Frank[1], Andreas Tittl[*,2] & Harald Giessen[*,1]

[†]These authors contributed equally to this work

[*]Corresponding Author

[1]4th Physics Institute, Research Center SCoPE, and Integrated Quantum Science and Technology Center, University of Stuttgart, Germany

[2]Chair in Hybrid Nanosystems, Nano-Institute Munich, Department of Physics, Ludwig-Maximilians Universität München, Germany

**Abstract**

Optical skyrmions are members of the emerging topological branch of solid-state physics and photonics, allowing for control over topological light textures through light-matter interactions. However, in nanophotonics their practical application has been severely limited by high inherent losses in plasmonic materials, resulting in the lack of tunability between different topological properties. Here, we utilize the strong dispersion of silicon carbide thin films to realize highly confined surface phonon-polariton skyrmion lattices, which we image via near-field microscopy. We experimentally demonstrate topological tuning between bubble- and Néel-type skyrmions, a unique advantage that polar dielectrics offer over most existing approaches. Changing the excitation wavelength by only 10% switches the skyrmion type, revealed by examination of the skyrmion number density contrast. Analysis of domain wall size and steepness in analogy to magnetic materials also confirms this transition.

Our results are a starting point to investigate other topological features in *phononic* systems such as merons, skyrmion bags, and other complex structured light fields. Furthermore, strong light-matter hybridization and nonlinear effects owing to anharmonicity of the phonons may be observed in the future, possibly leading towards the discovery of polaritonic skyrmion-skyrmion interactions and hence applications in topology-based information processing.

**Introduction**

Topological defects are characterized by their remarkable stability and robustness against small perturbations and continuous deformations of the lattice in which they reside. In recent years, they have become highly relevant in a plethora of fields in physics, ranging from cosmology[1] and hydrodynamics[2], to magnetism[3,4] and photonics[5–7]. Particularly in condensed matter physics, topological defects have recently found applications to study liquid crystals[8,9], reservoir computing[10], and twistronics[11].

A particular type of topological defect is the skyrmion. Originally introduced by Tony Skyrme in the context of particle physics[12], skyrmions are three-dimensional vector field configurations which twist in a continuous and non-trivial manner. These configurations are characterized by their topological charge, which represents the number of times the vector field wraps around a sphere[13]. Skyrmions have recently garnered attention in solid-state physics, particularly in the field of magnetism[14,15], where they arise through various properties such as pair-wise exchange interactions between neighboring magnetic spins and Dzyaloshinskii-Moriya interaction[16]. Skyrmions have also been explored in other systems, including elastic waves and hybrid platforms, emphasizing their potential for robust topological control[17–20].

In recent breakthroughs, the field of skyrmions has been extended to photonics[21,22], in which hexagonal lattices of optical skyrmions were generated and observed via interfering surface plasmon polaritons (SPPs) in thin gold films. Beyond their fundamental interest, optical skyrmions enable robust light–matter interactions, resilience to perturbations, and new opportunities for topological photonic functionalities[20,23,24]. The resulting field textures correspond to bubble-type skyrmions, which exhibit a vanishingly small, radially aligned in-plane field. Bubble-type skyrmions can be interpreted as a degeneracy of Bloch and Néel-type skyrmions known from magnetism[25]. In Néel-type skyrmions, the in-plane vector field points *radially*, resulting in a hedgehog-like configuration around the unit sphere. In contrast, Bloch-type skyrmions[26], exhibit a *tangential* rotation of the in-plane field, forming a vortex-like structure.

The ability to tune topological properties has been a long-standing goal in optics. Current approaches mostly focus on guided modes in metals via the excitation of surface plasmon polaritons to control different topological properties, such as orbital angular momentum (OAM)[27,28], but suffer from inherent limitations. For most metals, the highly dispersive region where enhanced tunability of the polariton wavelength could be achieved is usually dominated by high losses, which can substantially decrease the lifetime of guided modes, resulting in many topological properties not being realistically attainable.

Recent advancements have demonstrated the existence of phononic skyrmions in elastic wave systems, where the hybrid spin of elastic waves was used to generate and characterize topological properties at macroscopic, centimeter-length scales[17]. However, precise *optical* tuning of these topological features and their realization at nanometric scales has not been achieved.

In the following, we briefly outline the material platform used in this work, the connection to optical skyrmions previously observed in plasmonic systems, and how our work extends these earlier approaches by enabling tunability of their topological properties.

Silicon carbide (SiC) is a semiconductor that behaves like a metal within its Reststrahlen band (792–972 cm$^{-1}$)[29], located between its transverse and longitudinal optical modes[30,31], due to its negative dielectric constant ε in this spectral region. SiC is known for its extreme k-tuning over a narrow excitation wavelength range due to its strong sublinear dispersion within the Reststrahlen band[29,32–36]. Provided that in-plane momentum matching is ensured, photons can couple to this material, generating surface phonon polaritons (SPhPs), which are electromagnetic surface modes propagating along the interface, similar to surface plasmon polaritons (SPPs)[31]. However, unlike SPPs, where photons couple to the collective charge oscillations (plasmons) in metals, SPhPs involve photon coupling to lattice vibrations (phonons) in polar dielectrics.[37] Previous studies have demonstrated the potential of SPhPs in isotropic materials, as well as hyperbolic phonon polaritons (HPhPs) in anisotropic materials[37,38], such as hBN[39,40], for the generation of complex field configurations, such as deeply subwavelength polaritonic vortices[41–43].

Prior experimental and theoretical work has shown that *Bubble-type* optical skyrmion topologies can be realized via interfering SPPs. These topological textures possess a strong out-of-plane electric field, while the in-plane electric field almost fully vanishes. Tsesses et al.[22], who first measured these photonic skyrmions on gold surfaces, proposed the *theoretical* possibility of also generating *Néel-type* skyrmions by significantly increasing the confinement of evanescent surface waves; however, this could not be realized in gold due to high plasmonic losses within the regime of strong confinement.

Here we introduce the concept of *phonon-polaritonic* skyrmions in thin SiC membranes via interference of SPhPs. To achieve this, we use a 200 nm thick SiC membrane that is not only flat to reduce losses, but has the additional benefit of being thinner than the skin depth of the SPhP. This leads to hybridization of the top and bottom SPhP modes, enabling higher confinement[29]. Additional considerations regarding the dielectric function and thickness, as well as their influence on reflectance/transmittance, can be found in Supplementary Note 1.

This allows us to realize the proposal from Tsesses et al.[22] by utilizing the strong sublinear dispersion of SiC to measure and characterize the *transition from bubble- to Néel-type skyrmions* through minute changes in excitation wavelength. We directly image the out-of-plane electric field of phonon-polaritonic skyrmions with phase-resolved scattering scanning near-field optical microscopy (s-SNOM)[44], a method widely used for the characterization of resonant nanostructures[45,46] and polaritonic imaging[47,48]. In transmission-mode s-SNOM, a sharp tip is used to capture near-field optical signals and achieve resolutions far beyond the diffraction limit[49], resolving the subwavelength features necessary to characterize optical skyrmions. Our approach enables to experimentally tailor the steepness of domain walls present in skyrmion topologies, which delineate two domains of opposing field vectors within a single unit cell.

**Results**

The structures used to excite skyrmion lattices are depicted in **Fig. 1a**. The chromium ridges fabricated by electron beam lithography (see Materials and Methods) are used to excite surface phonon polaritons (SPhPs) with circularly polarized illumination at normal incidence. To form a skyrmion, the six SPhP waves have to interfere constructively at the center of the structure. In a perfect hexagonal arrangement, circularly polarized light launches the SPhPs generated at the different edges at different times, causing phase delays at the center. This is compensated by a hexagonal structure where each edge is shifted outwards by a multiple of the SPhP wavelength $\lambda_{SPhP}/6$, creating a spiral-like design. Consequently, a specific hexagonal structure must be created for each wavelength. A sketch of the structure is shown in **Fig. S3**.

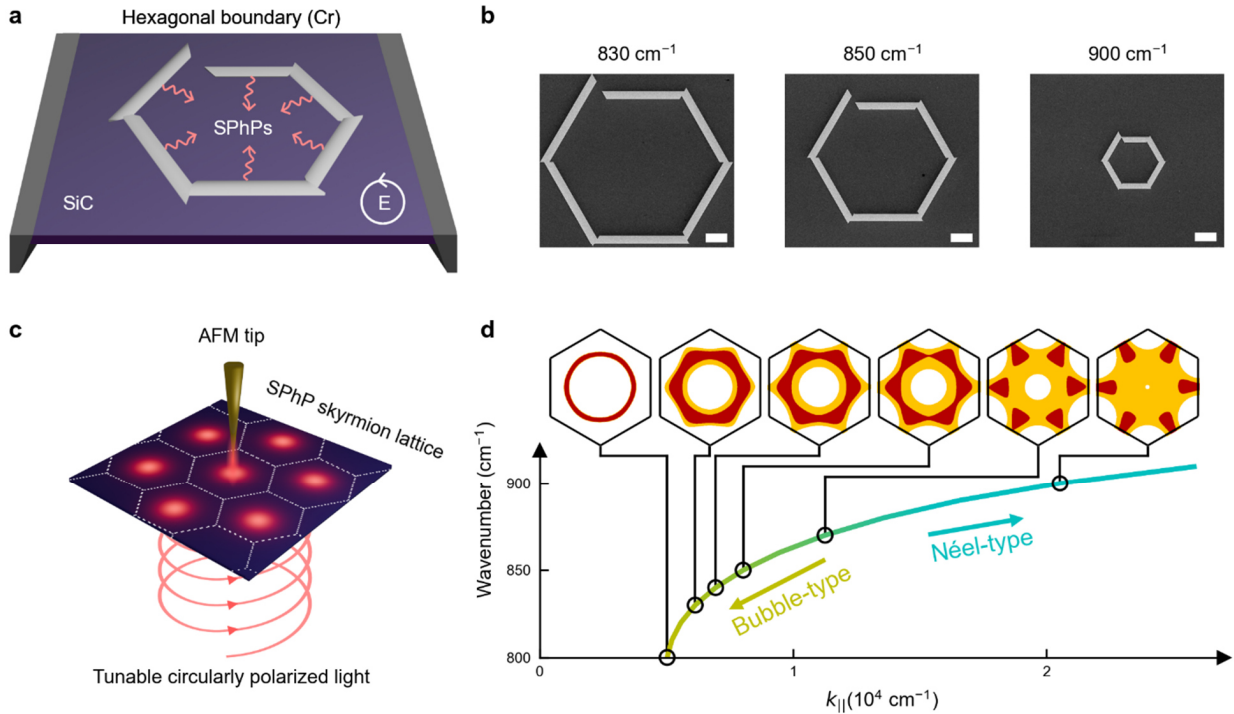

**Figure 1. SPhP Skyrmion lattice generation, tuning, and detection via near-field microscopy**. **a** For the formation of a phonon-polariton skyrmion lattice, six SPhP waves (as indicated by red arrows) are launched via hexagonal chromium strips on a 200 nm thick SiC membrane and interfere constructively at the center of each structure. The structures are illuminated with circularly polarized light at normal incidence. **b** SEM images of exemplary chromium structures designed for excitation at 830 cm$^{-1}$, 850 cm$^{-1}$, and 900 cm$^{-1}$. Scalebars 10 μm. The structures are placed in an s-SNOM operated in transmission mode for optical characterization (**c**), where circularly polarized laser pulses excite SPhP skyrmion lattices. While scanning across the SiC membrane, the s-SNOM tip interacts with the local near field and scatters the out-of-plane electric field to the far field (see Materials and Methods). Due to the strong dispersion of SiC (**d**), tuning of the wavenumber inside the Reststrahlen band leads to a drastic change of the in-plane momentum ($k_{||}$) of SPhPs, yielding a continuous transition from bubble- to Néel-type skyrmions, representing a meaningful topological property change captured in the field texture. In this graph, excitation regions of the different skyrmion types are color-coded. The corresponding skyrmion number densities are illustrated in a simplified representation (see **Fig. S12**), where high, intermediate, and low density regions are shown in brown, yellow, and white, respectively.

In **Fig. 1b** high resolution scanning electron microscope (SEM) images of the chromium structure for 830 cm$^{-1}$, 850 cm$^{-1}$, and 900 cm$^{-1}$ are displayed at normal incidence. These structures are placed in a transmission s-SNOM setup (**Fig. 1c**), where wavelength-tunable circularly polarized laser pulses are focused from below to excite phonon polaritonic skyrmion lattices, enabling measurements without interference from different excitation channels. The s-SNOM tip interacts with the local near field and scatters information to the far field, which is then recorded by an infrared detector. By scanning over the area of interest, an image of the local near field is created. The measured near-field signal is proportional to the out-of-plane electric field $E_z$, depicted in red for multiple skyrmions in **Fig. 1c**. By operating the tip in tapping mode to modulate the near-field signal with the tapping frequency and combining it with a modulated reference beam, we can extract both amplitude and phase while minimizing far-field contributions using a technique known as pseudo-heterodyne detection with higher-order demodulation[50]. The full setup can be found in **Fig. S4.**

A skyrmion is a topological field configuration which can be characterized by the skyrmion number density

$$s = \frac{1}{4\pi} \hat{e} \cdot \left(\frac{\partial \hat{e}}{\partial x} \times \frac{\partial \hat{e}}{\partial y}\right) \tag{1}$$

where $\hat{e} = E(r,t)/|E(r,t)|$ is the unit vector of the electric field[22]. The skyrmion number or winding number is calculated by $S = \int_\sigma s dA$, where $\sigma$ is the region in which the quasiparticle is defined. In our case of six interfering waves, a hexagonal skyrmion lattice is created with a single skyrmion in every unit cell. When integrating over the area of any unit cell, the result should be equal to 1. Although the skyrmion number remains constant, the distribution of the skyrmion density within the unit cell can be tuned by varying the electric field. The topology of skyrmions can be analyzed quantitatively using the skyrmion number density contrast (SNDC)[22]

$$\Psi = \frac{s_{max} + s_{min}}{s_{max} - s_{min}} \tag{2}$$

which is a function of the maximum and minimum value of the skyrmion number density *s* within the unit cell of a skyrmion and ranges from 1 to 0.5, enabling to determine the topology of optical skyrmions. We then tune the skyrmion number density by altering the wavenumber of the excitation laser in an area of strong dispersion (see **Supplementary Note 1**). More formal topological classifications based on winding numbers and related indices are discussed in the literature[51], whereas the tuning studied here occurs within a fixed skyrmion number and does not constitute a topological phase transition in the strict sense.

**Fig. 1d** illustrates how the in-plane wavevector changes with variations in the excitation wavenumber, calculated for a membrane thickness of 200 nm. With Gauss' law it is possible to relate the ratio of the in-plane and out-of-plane electric fields to the ratio of the corresponding wavevectors

$$\frac{E_\parallel}{E_z} = \frac{-k_z}{k_\parallel} \tag{3}$$

This formula not only helps us to reconstruct the 3-dimensional electric field from $E_z$ (which we obtain in s-SNOM measurements) but also allows us to simulate the skyrmion number density for different excitation wavenumbers. The panels at the top of **Fig. 1d** illustrate the skyrmion number density for different wavenumbers in a simplified representation. When the SPhP in-plane momentum increases, the skyrmion number density transitions from a ring or bubble shape to a gearwheel-like (Néel-type) shape, capturing a meaningful topological property change. The comparison between simulation and simplified representation can be found in the **Supplementary Fig. S12** and the skyrmion number density for selected values in **Fig. 2**. Additionally, the illustrations are linked to their position in the plot and the skyrmion number density contrast is color-coded.

Bloch-type skyrmions known from magnetism have so far not been generated using surface waves[52]. The Néel-type skyrmion exhibits a wide range of topological properties depending on the ratio between in-plane and out-of-plane electric field. When the in-plane electric field vanishes, it creates a degenerate case between Bloch- and Néel-type skyrmions, known as bubble-type. To differentiate between both cases quantitatively, we utilize the skyrmion number density contrast introduced by Tsesses et al.[22] Bubble-type skyrmions exhibit a skyrmion number density contrast close to 1, whereas Néel-type skyrmions can be characterized by a contrast of around 0.5.

In **Fig. 2,** we simulated a skyrmion lattice using Huygens waves (see **Supplementary Note 2**) for various wavelengths. The distance from the center to the closest ridge was set to three times the SPhP wavelength ($\lambda_{SPhP}$) (see **Supplementary Fig. S3**). This choice ensures constructive interference and results in a skyrmion lattice with 19 sites. To gain an intuitive understanding of the physics behind the skyrmion type transition, it is helpful to examine the corresponding 3D vector plot of the normalized electric field (**Figs. 2a-d**). It is evident that the transformation arises solely from the change in the ratio between the in-plane and out-of-plane electric fields.[22] To highlight the change of the in-plane electric field, a 2D cross section of the 3D vector plot is added.

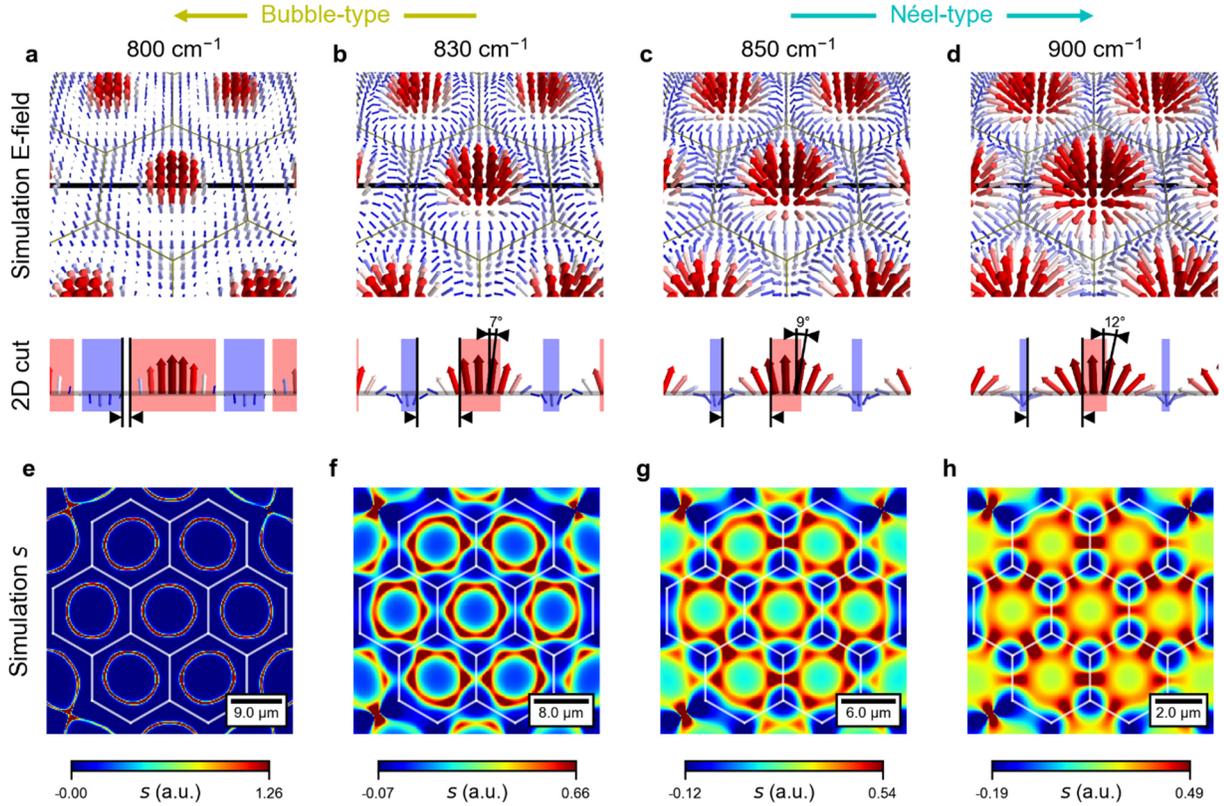

**Figure 2. Simulations of SPhP skyrmions via different excitation wavelengths. a-d** 3D vector plots of the simulated electric field. The grey line marks the cutting plane for the 2D cross section. The increasing strength of the in-plane electric field, which is visible from the orientation of the electric field vectors, leads to the continuous transition from bubble- to Néel-type skyrmions. For better visibility, the increase in polar angle $\theta$ and the domain wall size are highlighted. Domain walls are marked by white areas between the red (arrows up) and blue (arrows down) domains, indicated by triangular black arrows. **e-h** The Skyrmion number density $s$ corresponding to (a-d) exhibits a skyrmion lattice transitioning from bubble- to Néel-type. The white hexagonal grid delineates unit cells of the lattice.

The increasing strength of the in-plane electric field causes the polar angles $\theta = \tan^{-1}(E_{\parallel}/E_z)$ to rotate slower over a larger distance. For clarity, the polar angle is marked at an exemplary position (shifted by 7% of the unit cell width away from the center): it shifts from 0° at 800 cm$^{-1}$ to 7° at 830 cm$^{-1}$, 9° at 850 cm$^{-1}$ and 12° at 900 cm$^{-1}$. If we conceptualize the system as consisting of two distinct domains in analogy to magnetic skyrmions (electric field vectors pointing upwards and downwards, respectively, as indicated by red and blue areas in **Figs. 3a-d**.), this increase results in a wider domain wall between the domains. The figure also illustrates the different domains and their corresponding domain wall sizes.

The images in **Figs. 2e-h** illustrate a skyrmion lattice transitioning from bubble- to Néel-type; for better visibility, individual skyrmions are separated by a white hexagonal grid. The skyrmion number density is color coded and serves as a measure of the direction change of the electric field vectors. Therefore, it is not surprising that when the in-plane field is absent or very small, the value of the *s* changes instantaneously. This results in high skyrmion number density values in a small area, creating a bubble-type skyrmion, while the *s* remains near zero in all other areas. As the in-plane electric field increases, the transition of the electric field from up to down occurs gradually, leading to a smeared-out *s* and negative components around the edges of the unit cell. For a bigger in-plane electric field $s_{min}$ approaches $-s_{max}/3$, resulting in a SNDC of $\Psi = 0.5$. Hence, this yields a Néel-type skyrmion lattice.

In **Fig. 3**, we present experimental measurements of the continuous transition between bubble- and Néel-type topologies revealed via phase-resolved transmission-mode s-SNOM. For all measurements, the out-of-plane electric field $E_z$ exhibits seven skyrmions arranged hexagonally, as determined by the geometry of the hexagonal structures used to launch the SPhPs. Measurements near 800 cm$^{-1}$, close to the TO phonon frequency, were not experimentally feasible due to low signal-to-noise ratios, alignment challenges, and high reflectance arising from the large real part of the permittivity; we note that this frequency regime is already well-characterized in plasmonic systems. At 830 cm$^{-1}$ (**Fig. 3a**), the generated lattice strongly resembles a bubble-type skyrmion, which theoretically occurs precisely at the TO phonon frequency at around 792 cm$^{-1}$, when $\lambda_{SPhP} \approx \lambda_0$. Although the $E_z$ component appears qualitatively similar across different skyrmion types, the distinction between bubble- and Néel-type topologies becomes clear only when analyzing the full 3D vector field. This is particularly visible in the images plotting the polar angle $\theta$, which show a drastic transition from 0 to $\pi$ in the radial direction within each unit cell and thus exhibits sharp domain wall characteristics of this type of skyrmion. On the other hand, the radial transition of the polar angle between 0 and $\pi$ is much smoother in the case of 850 cm$^{-1}$ (**Fig. 3b**) and especially 900 cm$^{-1}$ (**Fig. 3c**), where $\lambda_{SPhP} < \lambda_0$, in good agreement with simulations shown in **Fig. 2**.

Additionally, the skyrmion number density $s$ calculated from the measured out-of-plane electric field displays different topological features depending on the incident wavelength. In good agreement with simulations in **Fig. 2f**, the skyrmion number density for 830 cm$^{-1}$ exhibits large values in a thin ring-shaped area, reminiscent of bubble-type structures, whereas the measurements at 850 cm$^{-1}$ and 900 cm$^{-1}$ resemble a topology close to the gearwheel-shaped Néel-type and agree well with simulations in **Figs. 2g, h**. The number inside each hexagon denotes the skyrmion number $S$ (calculated using eq. 1) and is relatively close to 1 for all measurements, thus confirming the high stability and robustness associated with optical skyrmions. This is also confirmed by the phase stability depicted in **Fig. S5,** together with the raw data, where the skyrmion number is calculated for various offsets added to the optical near-field phase ranging from $-\pi$ to $\pi$ and mostly takes on values of only 1 and -1, as expected from theory. Measurements of SPhP propagation lengths as well as data analysis methods including Fourier filtering are illustrated in **Figs. S6, S7, S8**. Thus, we have demonstrated the transition between bubble- and Néel-type optical skyrmion lattices, which is challenging to measure in most plasmonic materials such as gold due to strong plasmonic losses in regions where $E_\parallel \approx E_z$ and had, to our knowledge, not been achieved yet.

The definition of the skyrmion number density contrast as defined in eq. 2 is not well-suited for experimental data, as it relies on the minimum and maximum pixel values of the skyrmion number density $s$ within a unit cell of a skyrmion, which are often outliers. These outliers may arise from noise, local defects, or finite spatial resolution effects. As a result, the minimum and maximum values are not sufficiently robust to represent the characteristic contrast of the skyrmion texture. This effect is illustrated in Supplementary Fig. S9. Therefore, for both, the measured and simulated data, we employed a slightly adjusted method, using the 5th and 95th percentiles instead of the maximum and minimum values. This approach significantly reduces the influence of outliers while preserving the physically relevant contrast.

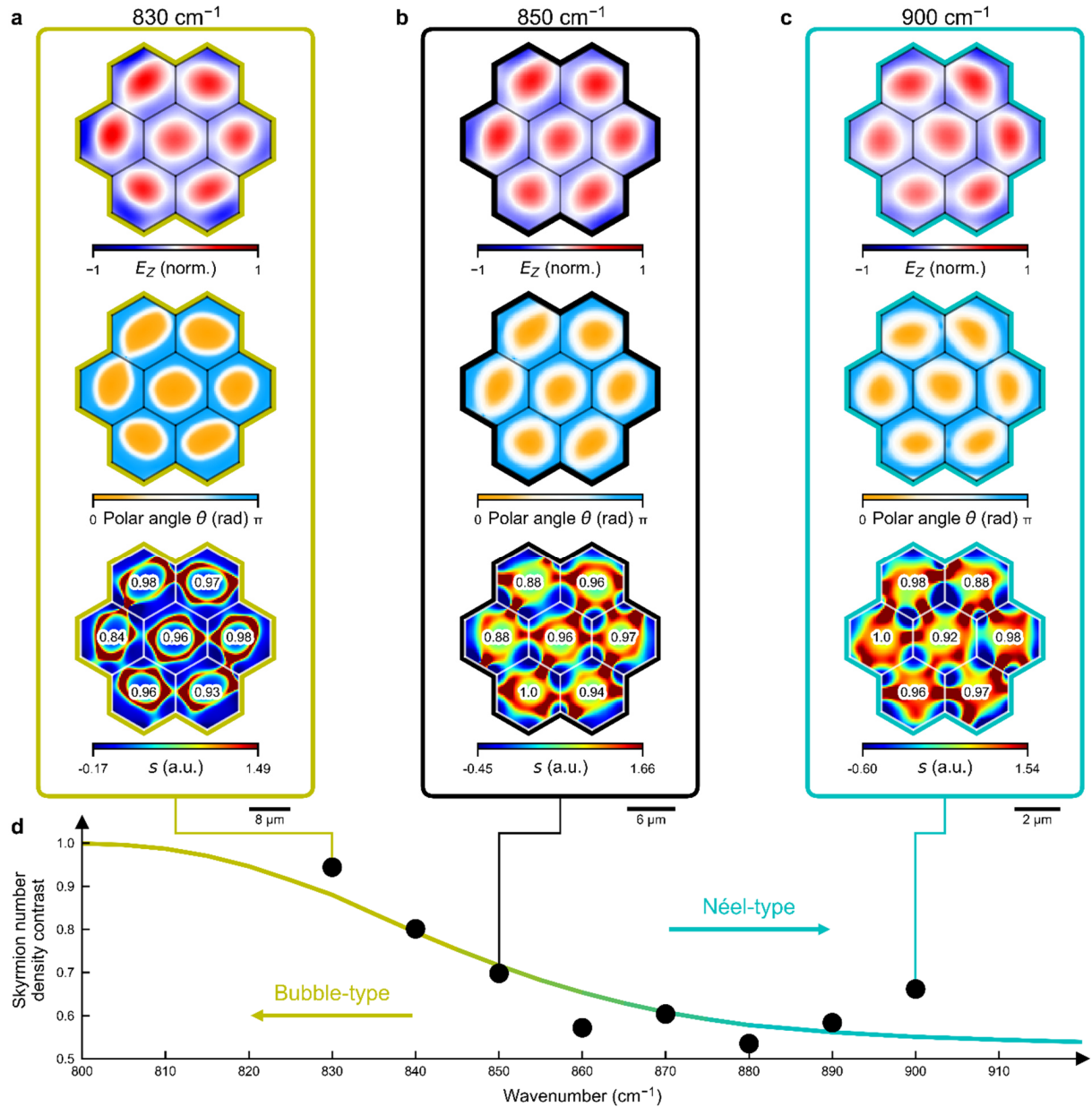

**Figure 3. Experimental measurements of SPhP skyrmion lattice transition from bubble- to Néel-type. a-c** Transition from bubble- to Néel-type skyrmions, depicting the real part of the out-of-plane electric field $E_z$, polar angle $\theta$ as well as the skyrmion number density measured at 830 cm$^{-1}$ (**a**), 850 cm$^{-1}$, (**b**) and 900 cm$^{-1}$ (**c**). As expected from theory, the real part of $E_z$ remains independent of the excitation wavelength. On the other hand, the polar angle $\theta$ displays clearly delineated domains for bubble-type skyrmions and smeared-out domains for Néel-type skyrmions within each unit cell. The skyrmion number density $s$ illustrates the continuous transition from bubble-shaped to gearwheel-like (Néel-type) configurations, closely matching the simulations in **Fig. 2e-h**. The skyrmion number for each skyrmion in the lattice is approximately 1. (**d**) Tuning the skyrmion number density contrast from bubble- to Néel-type. The coloured line represents the simulated values, while the points represent the experimental data.

**Fig. 3d** illustrates the calculated SNDC versus incident light frequency. The colored line represents the simulated values with the adjusted SNDC described above, while the dots illustrate measured values, demonstrating the continuous transition from Néel- to bubble-type skyrmions with six intermediate steps. While fabrication inaccuracies, sample roughness, and the spectral width of the laser may have a potential impact on the quality of measurement, our results agree reasonably well with simulated SNDC values.

As additional figures of merit (FOMs), to characterize the topological states of optical skyrmion lattices, we introduce the inverse domain wall steepness and the domain wall size. Inspired by magnetic materials, these metrics focus on the relationship between the in-plane and out-of-plane components of the electric field, which differentiates bubble- from Néel-type skyrmions. Both FOMs are derived from the polar angle $\theta$, capturing key aspects of the skyrmion transition.

To calculate the inverse domain wall steepness (DWS), we sort the data based on the distance from the center of each skyrmion and fit a suited analytical function to it (see **Supplementary Fig. S9**). The derivative of the fitted function with respect to the distance $r$ from the center provides the steepness, and since the steepness can approach infinity at certain wavelengths (e.g., around 800 cm$^{-1}$), we consider its inverse to avoid impractical values for the FOM (see **Supplementary Fig. S10**).

$$DWS = \left(\frac{\partial \theta}{\pi \cdot \partial r}\right)^{-1} \tag{4}$$

Furthermore, the domain wall size is determined by measuring the distance between the 10% and 90% thresholds of the polar angle's minimal (0°) and maximal (180°) values. This approach is intuitive, easy to visualize, and provides physically meaningful data, though the choice of thresholds is somewhat arbitrary.

Both FOMs highlight the smoother change in polar angle for Néel-type skyrmions and the abrupt shift for bubble-type skyrmions. As depicted in **Fig. 4**, the simulated curves effectively distinguish between these topologies, and the inverse domain wall steepness aligns better with the simulations, attributed to the reduced impact of outliers in the measured data. This demonstrates that inverse domain wall steepness and domain wall size capture the essential topological features governed by the shape of the electric field. The individual data points as well as the fitted functions and the derived domain wall steepness can be found in Figs. **S9** and **S10**.

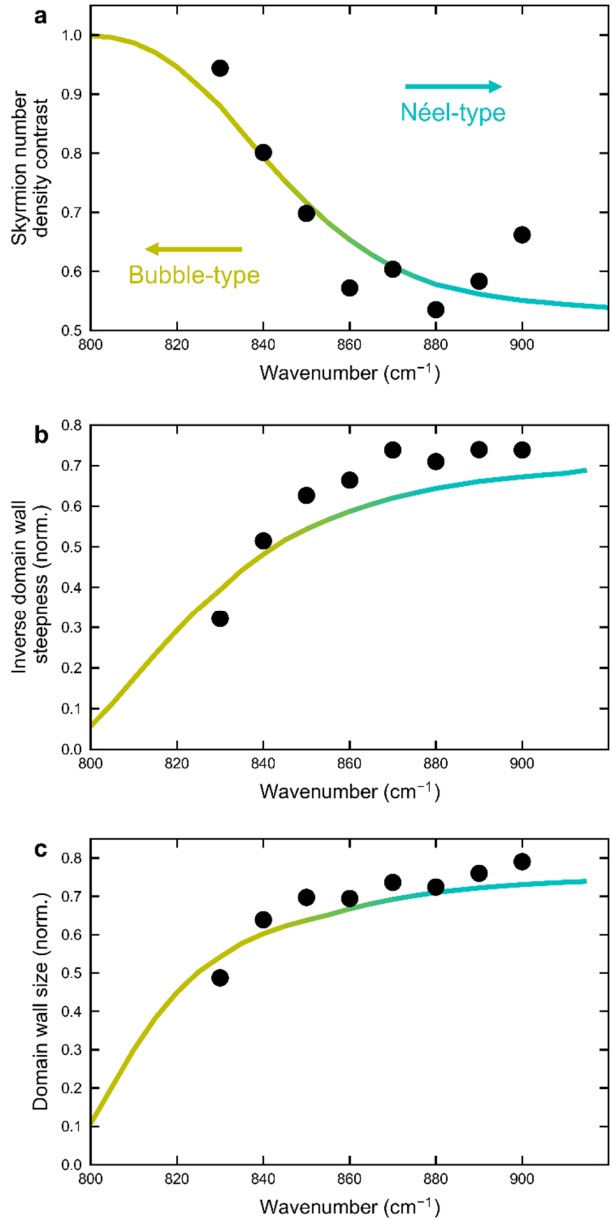

**Figure 4. Comparison of the figure of merit (FOM) for skyrmion number density contrast, inverse domain wall steepness, and domain wall size, to investigate the transition from bubble- to Néel-type skyrmions.** The lines represent simulated values, while the points indicate measured data, with normalization such that the unit cell length is set to 1 in all cases. **a** Skyrmion number density contrast using the previously applied method. **b**, **c** Polar angle-dependent FOM for domain wall steepness and domain wall size, respectively, capture the key aspects of the continuous transition from bubble- to Néel-type skyrmions. The experimental data shows better agreement with simulations, displaying the robustness of the FOMs introduced here against outliers in the data.

**Discussion**

We introduced the concept of phonon-polaritonic skyrmions in polar dielectrics. Through the interference of SPhPs in carefully designed structures, we provided both theoretical and experimental observations of skyrmion lattices free-standing thin SiC membranes. Additionally, this work provides the first clear measurements of SPhP Néel-type skyrmions, which are characterized by smooth boundary walls, overcoming previous limitations imposed by plasmonic systems. Exploiting the strongly dispersive nature of SiC, we are able to tune across a large range of topological properties and domain-wall configurations , visualizing the continuous transition between the previously well-studied bubble-type and the now available Néel-type optical skyrmion. Our system uniquely allows for the precise tuning between bubble- and Néel-type optical skyrmions at the nanometer scale using phonon-polaritonic guided modes.

Future research can make use of these findings by exploring the use of radially polarized light in order to potentially generate a wide range of different topologies within the same structure. This increase in flexibility could increase the likelihood for skyrmions to be applied for direct on-chip computing purposes in the future[53], possibly storing information in the Chern number. In addition, small antennas instead of continuous structures could be employed to selectively shape and steer the local field distribution, offering new avenues for generating complex topological configurations and enhancing device versatility[54]. Twisting bi-hexagonal structures relative to each other can create new topological configurations, paving the way for phonon polaritonic twistronics and the transition of skyrmion bags from bubble to Néel-type[55–57].

Finally, the use of other materials supporting guided modes, such as various 2D-materials, could foster developments in this field and enable a new host of topological light textures. In particular, actively modulating material properties such as carrier density in, e.g., graphene via electrical or optical switching may allow for rapid transition between various skyrmion topologies, tuning between skyrmion and meron topologies within the same structure[58]. Our discovery might also be utilized in structured light-matter interaction, utilizing phonons in the mid-IR spectral region.

In general, the unique properties of phonon polaritonic skyrmions have the potential to be harnessed for future applications in spin-optics, imaging, and topological and quantum technologies, making them ideal candidates for next-generation photonic and phononic devices. Furthermore, the inherent nonlinearity due to anharmonic lattice potentials in phonons opens up a possible road towards the long sought after polaritonic skyrmion-skyrmion interaction. This might enable topology-based information processing in such phononic systems.

**Materials and methods**

Optical Near-field Measurements

Experimental measurements were conducted using a commercial s-SNOM system (neaSCOPE from attocube systems, Haar, Germany) with the transmission module in pseudo-heterodyne detection scheme. The laser source is a fs OPO (Stuttgart Instruments Alpha, Stuttgart, Germany) with a 1030 nm pump laser and tunable MIR output obtained through DFG generation in a nonlinear crystal ranging from $\lambda = 4.3$ μm to $\lambda = 16$ μm. A grating monochromator is placed between laser source and microscope to limit the spectral width of the beam to 10 cm$^{-1}$. The power used for measurements is around 1 mW.

Upon entering the microscope, the beam is split by a beam splitter. One part of the beam traverses a $CdGa_2S_4$ quarter-wave plate to generate circularly polarized light before being loosely focused onto the SiC membrane via a parabolic mirror. To excite optical skyrmions and minimize distortions stemming from uneven illumination, the beam should illuminate the structure homogenously. This can be achieved through manually defocusing the beam by changing the height of the bottom parabolic mirror. As we verified experimentally, the resulting beam spot is larger than the maximum scanning area (>100x100 μm$^2$). An additional iris was placed after the monochromator to cut the beam tails away and obtain a wavefront which resembles a plane wave more closely. Measurements conducted to verify these requirements are shown in **Fig. S11**.

The probing s-SNOM tip operates in tapping mode at frequencies of 200-300 kHz (Arrow-NCPt tips, $r = 25$ nm) and at tapping amplitudes of 70-90 nm. The backscattered light from the tip is collected by a second parabolic mirror placed above the tip. The other part of the beam is modulated through a vibrating mirror with a wavelength-dependent mirror amplitude to decouple optical phase ($\varphi_n$) and amplitude ($s_n$) and then recombined through a beam splitter with the signal beam. The interference between signal and modulated reference beam is detected by a liquid nitrogen cooled MCT-detector and demodulated based on higher orders of the tip frequency ($n>1$) to eliminate background signal. In our normal-incidence transmission s-SNOM geometry, the detected near-field signal is dominated by edge-launched SPhPs, while tip-induced launching is negligible and does not contribute measurably to the data.[29] The main reason for this mechanism is that in transmission s-SNOM, the incident light couples to the sample from below, thus exciting SPhPs before reaching the tip. Therefore, the tip acts as a passive scatterer in transmission mode, as opposed to reflection mode, where the beam is focused directly onto the tip before passing through the sample.

Fabrication

The free-standing 200 nm SiC film suspended on Si frame was purchased from Silson Ltd. The chromium ridges were fabricated following the standard top-down fabrication procedure. First, the masks were created on spin-coated 950 K polymethyl methacrylate with electron-beam lithography and subsequently metalized with electron-beam deposition. The final thickness of the chromium was set to 30 nm, and the structures were realized by lifting off the polymer masks in acetone.

**Acknowledgments**

Funding: The authors acknowledge support from the ERC (Complexplas, 3DPrintedoptics), DFG (SPP1391 Ultrafast Nanooptics, CRC 1242 "Non-Equilibrium Dynamics of Condensed Matter in the Time Domain" project no. 278162697-SFB 1242, Germany's Excellence Strategy EXC 2089/1–390776260, Emmy Noether Program TI 1063/1), BMBF (Printoptics), BW Stiftung (Spitzenforschung, Opterial), Carl-Zeiss Stiftung and from the DFG (GRK2642 Photonic Quantum Engineers). Funded by the European Union (ERC, METANEXT, 101078018 and EIC, NEHO, 101046329). Views and opinions expressed are however those of the author(s) only and do not necessarily reflect those of the European Union, the European Research Council Executive Agency, or the European Innovation Council and SMEs Executive Agency (EISMEA). Neither the European Union nor the granting authority can be held responsible for them.

**Contributions**

H.G. conceived the idea of phonon-polaritonic skyrmions; F.M. conceptualized the experiment and designed the samples with L.N.; L.N. fabricated the sample; E.B. and F.M. conducted experiments and analyzed the results with assistance from T.G., A.M., B.F.,A.T.; F.M. and J.S. did the skyrmion number analysis with assistance from E.B.; J.S. developed the theoretical framework and envisioned the skyrmion transition in SiC; F.M. and E.B. wrote the manuscript; all authors provided valuable advice and helped refine the manuscript; B.F., A.T. and H.G. managed various aspects of the project.

**Data availability**

All data required to support the conclusions presented in this paper are included either in the main text or in the supplementary materials. Additional datasets and analysis programs can be obtained from the authors upon reasonable request.

**Conflict of interest**

We report the following conflict of interest: T. Gölz received financial support for his PhD thesis from attocube systems AG.